\documentclass[12pt]{iopart}

\usepackage{iopams}

\newcommand{\Z}{\mathbb{Z}}
\newcommand{\N}{\mathbb{N}}
\newcommand{\beq}{\begin{equation}}
\newcommand{\eeq}{\end{equation}}

\newcommand{\xx}{\mathbf{x}}
\newcommand{\eps}{\varepsilon}
\newcommand{\mafrac}{\sqrt{\frac{8 \pi}{3}}}

\begin{document}

\title{MG13 Proceedings: A lattice Universe as a toy-model for inhomogeneous cosmology}

\author{J. -P. Bruneton$^1$}

\address{$^1$Namur Center for Complex systems (naXys), University of Namur, Belgium}
\ead{jpbr@math.fundp.ac.be}

\begin{abstract}
We briefly report on a previously found new, approximate, solution to Einstein field equations, describing a cubic lattice of spherical masses. This model mimics in a satisfactory way a Universe which can be strongly inhomogeneous at small scales, but quite homogeneous at large ones. As a consequence of field equations, the lattice Universe is found to expand or contract  in the same way as the solution of a  Friedmann Universe filled with dust having the same average density. The study of observables indicates however the possible  existence of a fitting problem, i.e. the fact that the Friedmann model obtained from past-lightcone observables does not match with the one obtained by smoothing the matter content of the Universe.
\end{abstract}


\section{A toy-model for inhomogeneous cosmology}\label{aba:sec1}
As a toy-model of inhomogeneous Universe, we study a lattice Universe of, e.g., galaxy-like objects. 
The parameters of the lattice are the comoving size $L$ of the cells, the mass $M$ inside each cell, and $\eta$ the (comoving) ``size'' of the masses themselves. This means that we choose to describe the source term by a three dimensional comb of Gaussians of width $\eta$. Two dimensionless ratios then govern the physics of the model: the typical depth of gravitational potential, $G M/ L c^2$, and the degree of inhomogeneity of the lattice Universe, $\eta/L$. When $\eta/L$ is of order $1$, the lattice Universe is nearly homogeneous at all scales with very low density contrasts while small values $\eta/L \ll1$ corresponds to a Universe essentially made of voids and of small regions of very dense matter. Relevant parameters, phenomenologically speaking, are typically galaxy-sized masses, $M \sim 10^{11}$ Solar masses, and $L \sim 1 Mpc$, for which the ratio $G M/Lc^2$ is very small, of order $10^{-8}$.\\
\\
This last point suggests to try and solve the Einstein equations perturbatively, as a power series in $GM/Lc^2$. Doing so, Einstein equations are effectively linearized up to order three in the parameter $\sqrt{M/L}$ which turns out to be the relevant quantity for Taylor expanding the metric of spacetime and the stress-energy tensor \cite{Bruneton:2012cg}. Moreover, one can take advantage of the periodicity of the lattice, and thus look for a periodic solution. For simplicity, we have assumed flat space with no cosmological constant. The solution then reads \cite{Bruneton:2012cg}, in synchroneous gauge, $g_{00}=-1$, $g_{0i}=0$, and
\begin{eqnarray}
\label{mainresultmetric}
g_{ij}= \delta_{ij} \left[1+ 2\varepsilon\sqrt{\frac{GM}{Lc^2}}  \mafrac \frac{c t}{L} +\frac{2GM}{L c^2}\left(f_{\eta}(\mathbf{x})+ \frac{2 \pi c^2 t^2}{3 L^2} \right) \right.
\nonumber\\
\left. +  2 \left(\frac{GM}{Lc^2}\right)^{3/2}\left( 2 \varepsilon  \frac{c t}{L} \mafrac  f_{\eta}(\mathbf{x})  - \frac{2 \pi \varepsilon}{9}\mafrac \frac{c^3 t^3}{L^3}\right) \right] \nonumber\\
+\frac{G M}{L c^2} c^2 t^2\partial_{ij}^2 f_{\eta}(\mathbf{x}) 
+ \left(\frac{GM}{Lc^2}\right)^{3/2} \eps\sqrt{\frac{8 \pi}{3}} \frac{ c^3 t^3}{3 L} \partial_{ij}f_{\eta}(\mathbf{x}) +\mathcal{O}\left(\frac{M^2}{L^2}\right),
\end{eqnarray}
where $\varepsilon = \pm 1$ corresponds to an expanding or contracting lattice, and the function $f_{\eta}$ which accounts for the anisotropies of the gravitational field reads:
\begin{equation}
f_{\eta}(\xx)=\frac{8}{\pi}\sum_{(n,p,q)\in\N^{3}_{*}}\frac{e^{-\frac{\pi^{2} (n^2+p^2+q^2)\eta^{2}}{L^{2}}}\cos\left(\frac{2\pi}{L}nx\right)\cos\left(\frac{2\pi}{L}py\right)\cos\left(\frac{2\pi}{L}qz\right)}{n^{2}+p^{2}+q^{2}}. \label{feta}
\end{equation}
This metric solves Einstein equations $G_{\mu\nu} = 8\pi G/c^4 T_{\mu\nu}$  with no cosmological constant, and for a source term representing the lattice up to order $(M/L)^{3/2}$, namely a stress energy tensor whose only non-vanishing component reads
\beq
T_{00} = \frac{M}{L^3} \left( 1-\frac{1}{2}\sqrt{\frac{M}{L}} \sum_{i} h_{ii}^{(1)}\right) \sum_{\mathbf{n}\in\Z^3}e^{\frac{2\pi}{L}i\mathbf{n}.\xx-\frac{\pi^{2}|\mathbf{n}|^{2}\eta^{2}}{L^{2}}}+\mathcal{O}\left(\frac{M^2}{L^2}\right),
\eeq
where $g_{\mu\nu}=\eta_{\mu\nu} + \sqrt{\frac{M}{L}} h^{(1)}_{\mu\nu} +\mathcal{O}(M/L)$ defines $h_{\mu\nu}^{(1)}$ via Eq.~(\ref{mainresultmetric}). The solution is typically valid for a limited range in time and space $\Delta t$ and $\Delta x$, with $\Delta x \sim \Delta t \ll L \sqrt{L/M} \sim 10 Gpc$. \\
\\
The metric enables one to compute the effective scale factor for such a Universe, and thus an effective Hubble parameter. Up to the order of approximation considered, one finds \cite{Bruneton:2012cg}
\beq
\label{Hfinal}
H(t) =  H_0 -\frac{3}{2} H_0^2 t+\frac{9}{4} H_0^3 t^2 +\mathcal{O}(H_0^4),
\eeq
where the Hubble parameter at time $t=0$ is identified with the comoving average density $H_0 = \eps \mafrac \sqrt{\frac{G M}{L^3}}$. 
This corresponds to a flat Friedmann-Lemaitre model with equation of state $w=0$. Thus, the model with discrete masses on a cubic lattice is identical to a Friedmann model with dust, with the corresponding energy density. This means that one cannot distinguish the distribution of mass (localized or smeared in an homogeneous manner) from purely kinematical considerations. In the language introduced in \cite{Kolb:2009rp}, the model exhibits no strong backreaction, ie. the scale factor of the averaged Universe is identical to the average of the scale factor, up to the order of approximation considered, in accordance with other studies in similar contexts, see \cite{Yoo:2012jz, Clifton:2012qh}.

\section{Observables and the fitting problem}\label{aba:sec2}
This sort of commutativity property is however broken at the level of observables. In \cite{Bruneton:2012ru}, we have studied the propagation of bundle of light rays in the spacetime given by the previous solution. By solving the geodesic equation, we have computed, up to order $GM/Lc^2$, the redshift as a function of the affine parameter $\lambda$, while solving perturbatively the Sachs equations governing the expansion and the shear of the bundle, we have derived an analytical expression for the angular distance as a function of $\lambda$. Combining both, this leads to distances as a function of redshift.\\
\\
Doing so, one finds that observables are given by their standard expression in the analog Friedmann model with the average density, plus non-trivial corrections that need not to stay negligible in general\cite{Bruneton:2012ru}. This indicates that a weak backreaction\cite{Kolb:2009rp}, or a fitting problem\cite{Ell84}, is present in the model. By carefully analysing the analytical expressions found, we have been able to establish a quantitative criterion, connecting the three length scales of the problem, that decides whether deviations to the laws of homogeneous cosmology are to stay small. More precisely, we showed that in a Universe where masses are smeared enough, namely, satisfying
\beq
\label{mainbound}
\frac{M}{L} \ll \mathcal{O}(1) \times \left(\frac{\eta}{L}\right)^4,
\eeq
the perturbations to Friedmann's expression for the observables remain small. A violation of this bound, on the other hand, ie. when objects are too compact, or the Universe too inhomogeneous, implies a breakdown of the perturbative expansion, as far as the calculation of observables are concerned, although the pertubative scheme remains relevant in order to estimate the spacetime geometry. The next step is therefore to study more precisely the behavior of observables by solving non-perturbatively the coupled system of Sachs equations in the perturbed metric Eq.~(\ref{mainresultmetric}). Indeed, the $\sqrt{M/L}$ perturbative solution to Sachs equations cannot be trusted anymore in this case, because it decouples the shear from the expansion of the bundle of light rays, whereas the shear is expected to be large, and should thus play an important role. This is work in progress.\\
\\
In conclusion, we have found a realistic model for inhomogeneous cosmology that exhibits no strong backreaction, but weak backreaction whose amplitude is linked to the Weyl tensor sourcing for the shear, and therefore also linked to the ``compactness'' of the sources, or, if one wishes, to the degree of inhomogeneity of the Universe.\\
\\

\bibliographystyle{ws-procs975x65}
\bibliography{Torus_biblio}

\end{document}